\documentclass[aps,amssymb,floatfix,prd,amsmath,preprintnumbers]{revtex4}
\usepackage{epstopdf}
\usepackage{capt-of}
\usepackage{graphicx}  
\usepackage{dcolumn}   
\usepackage{bm}
\usepackage{hyperref}
\usepackage{ulem}
\usepackage{xcolor}
\begin{document}
\input epsf.tex
\title{Viscous fluid accelerating model in modified gravity}
\author{Sankarsan Tarai,\footnote{Centre of High Energy and Condensed Matter Physics, Department of Physics, Utkal University, Vani Vihar, Bhubaneswar, India-751004 E-mail: tsankarsan87@gmail.com}, Pratik P Ray \footnote{Department of Mathematics (SSL), Vellore Institute of Technology-Andhra Pradesh University, Andhra Pradesh - 522237, India, E-mail: pratik.chika9876@gmail.com} B. Mishra, \footnote{Department of Mathematics, Birla Institute of Technology and Science-Pilani, Hyderabad Campus, Hyderabad-500078, India, E-mail:bivu@hyderabad.bits-pilani.ac.in}, S.K. Tripathy \footnote{Department of Physics, Indira Gandhi Institute of Technology, Sarang, Dhenkanal, Odisha 759146, India, E-mail:tripathy\_sunil@rediffmail.com }
}
\affiliation{ }

\begin{abstract}
In this paper, we have investigated the late time cosmic acceleration issue in the context of $f(R,T)$ gravity. The matter field is considered to be that of viscous fluid. The model has been framed as a mathematical formalism and the effect of viscous fluid on the cosmic expansion has been shown. The equation of state parameter indicates the quintessence behaviour of the Universe at late time. The theoretical results obtained here shows its alignment with the cosmological observations result.
\end{abstract}
\maketitle

\section{Introduction}
 
The need of an alternative theory to Einstein's General Relativity (GR) or any other alternative theory had become inevitable, when the existing theories unable to answer some key issues on the gravitational phenomena. Some theories are more phenomenological and others motivated by theoretical and observational results. In a similar note, an alternative to GR has become inevitable when several cosmological observations claim the accelerated expansion of the Universe \cite{Reiss98,Reiss99,Spergel03,Abaz24,Pope24}. In addition, several observations have indicated the presence of some kind of aether or dark energy in the Universe. GR and other alternative theories of gravity could not make a satisfactory answer to this claim. Therefore, researchers motivated to find an alternative gravity theory to get an explanation pertaining to the present astrophysical observations. The search for a new gravity has not been confined to the reason of presence of dark energy, which is the main part of the energy budget of the Universe, but also for the huge amounts of unseen matter. It is a belief that this unseen matter may give some explanation to the formation of structure, gravitational lensing and the rotation curves of galaxies. Of course the study of accelerated expansion of the Universe can be performed with the unknown form of matter and energy driven through negative pressure, but the change in the geometric part of gravitational sector of GR received a lot of attention. \\ 

Modified theory of gravity is widely discussed to study the gravitational interaction\cite{Noji00, Noji11, Capo}. They are based on modification and enlargements of the Einstein theory. This can be obtained by adding higher order curvature invariants and minimally or non minimally coupled scalar fields into dynamics. Indeed, it provides a very natural unification of the early-time inflation and late-time acceleration in evolution of the Universe. The effective dark energy dominance may be assisted by the modification of gravity. Hence, it is believed that the coincidence problem can be solved by the expansion behaviour of the Universe. In the absence of fundamental quantum gravity, the phenomenological approach to modify the gravity theory by hand and then by complying the modified theory with observational data and data from local tests. The modified gravity approach possesses attractive features towards the application of accelerating Universe. It provides several natural gravitational alternatives for dark energy and dark matter. In fact Bamba et al. \cite{Bamba12} have given a clear description of Dark energy cosmology, its equivalent description through different theoretical models. The modified theory is able to describe the evolution journey from early time inflation and late time acceleration. The acceleration is nicely explained by different role of gravitational sub-dominant terms, such as $\frac{1}{R}$, which might become relevant at small and at large curvature. Also, this theory acts as the basis for cosmological effects such as the galaxies rotation curves. Modified gravity, describes the transition of Universe from non-phantom phase to phantom one without the help of exotic matter. Due to this transient nature, no future Big Rip is usually expected in the evolutionary scenario. This gravitational theory justifies the transition from early deceleration to late time acceleration of the Universe. Though, the constraints from solar system tests are quite stringent, modified gravity theories can take viable approach in order to precisely check the solar system tests.\\

Among the different suggested modifications to GR, the $f(R)$ gravity theory is a popular geometrically modified gravity that generalizes GR. However, Harko et al.\cite{Harko11} extended the $f(R)$ gravity by introducing the trace of energy momentum tensor $T$ in the action, known as $f(R,T)$ gravity. Many investigations are surfaced to find the cosmic nature of the Universe particularly on the accelerating Universe in $f(R,T)$ gravity. Under this gravity, Das et al. \cite{Das16} have obtained the solutions that describes the interior of a compact star whereas Deb et al. \cite{Deb18} have presented the spherically symmetric strange star solution. Fisher and Carlson \cite{Fisher18} have re-examined $f(R,T)$ gravity and argued that the separable trace term should be included in the matter Lagrangian. Shabani and Ziaie \cite{Shabani18} have studied the classical bouncing solutions in FRW background and Tripathy et al. \cite{Tripathy19} have presented the bouncing solution with exponential and power law cosmology. Barbar et al. \cite{Barbar20} have analysed the viability of bouncing cosmology by incorporating the square of energy momentum tensor. Wu et al. \cite{Wu18} have obtained the generalized Friedmann equations and studied its cosmological implications. Elizalde and Khurshudyan \cite{Elizalde19} have investigated static wormhole cosmological model in $f(R,T)$ gravity. In this gravity,  Khan et al. have studied the gravitational collapse of perfect fluid in a spherically symmetric space-time \cite{Khan18} and effects of electromagnetic field on gravitational collapse \cite{Khan19}. Tripathy and Mishra \cite{Tripathy20} have studied the phantom cosmology whereas Mishra and Tripathy \cite{Mishra20} investigated the little rip and hyperbolic form of scale factor in $f(R,T)$ gravity. Triapthy et al. \cite{Tripathy2020} have constructed some cosmic transit model in $f(R,T)$ gravity theory and studied their cosmographic aspects . Saridakis et al. \cite{Saridakis20} have investigated the cosmological implications of Myrzakulov $f(R,T)$ gravity. Several cosmological models have been constructed in $f(R,T)$ gravity to study different aspects of the late time cosmic phenomena.\\

In this context, the action of $f(R,T)$ gravity becomes,

\begin{equation} \label{eq:1}
S=\frac{1}{16\pi} \int f(R,T)\sqrt{-g}d^4x+\mathcal{L}_m \sqrt{-g} d^4x,
\end{equation} 
where $f(R,T)$ is an arbitrary function of Ricci scalar $R$ and $T=T_{ij}g^{ij}$; $T_{ij}$ is the energy momentum tensor. The matter Lagrangian,  $\mathcal{L}_m=-p$. The Lagrangian density of matter field depends only on the metric tensor component $g_{ij}$ and not on its derivatives. So, the stress energy tensor of matter can be, 

\begin{equation} \label{eq:3}
T_{ij}= g_{ij} \mathcal{L}_m-2\frac{\partial \mathcal{L}_m}{\partial g^{ij}}.
\end{equation}

By varying the modified four-dimensional Einstein-Hilbert action \eqref{eq:1} with respect to the metric tensor components $g^{ij}$, the modified field equations are obtained where we chose the algebraic function $f(R,T)$  as a sum of two independent functions $f(R,T)=f_1(R)+f_2(T)$, where $f_1(R)$ and $f_2(T)$ are respectively some  functions of curvature $R$ and trace $T$.\\

However, in order to derive a complete field equation, a matter source must be fitted to the matter Lagrangian that ultimately contributes to the stress energy tensor. In the early Universe, viscosity may originate due to processes like decoupling of matter from radiation in the recombination era, collision of particles involving gravitons or during the formation of galaxies \cite{Barrow1977}. Earlier attempts \cite{Pavon1993, Padmanabhan87} in this area even predict the late acceleration of the Universe expansion. The dissipative effect in the fluid is mostly due to shear and bulk viscosity characterized by shear viscosity parameter $\eta$ and bulk viscosity parameter $\zeta$. In the FLRW background the shear viscosity disappears and the realistic model can be obtained with bulk viscosity.  The effect of this can be assessed by adding the correction term as, $\bar{p}=p-3\zeta H$. This composite formula motivates the trial on the connection between bulk viscosity of the cosmic fluid and dark components. In this paper, we discuss the unified model considering $f(R,T)$ gravity with viscous fluid. Brevik et al. \cite{Brevik17b} have shown the significance of viscous fluid in the early and late Universe. Noureen et al. \cite{Noureen} have studied the evolution of spherically symmetric charged anisotropic viscous fluids in $f(R,T)$ gravity. Singh and Kumar \cite{CP} have obtained the exact non singular solutions with non-viscous and viscous fluids in the frame of $f(R,T)$ gravity. Ayugn \cite{S} has studied homogeneous and anisotropic Marder space-time with bulk viscous matter distribution solutions in $f(R,T)$ gravity. Azmat et al. \cite{H} have analyzed the role of shear viscosity and pressure anisotropy on dynamics of cylindrical system using perturbative approach in $f(R,T)$ gravity. Prasad et al. \cite{RP} investigated the bulk viscous accelerating universe compiling with observational Hubble data, the baryon acoustic oscillation data and SNLa data in $f(R,T)$ gravity. Debnath studied a bulk viscous cosmological model in $f(R,T)$ gravity theory \cite{Debnath2019}. In a recent work, Odintsov et al. have tested the equation of state for viscous dark energy \cite{Odintsov2020}.\\
 
The field equations of $f(R,T)$ gravity can be obtained as \cite{Harko11},

\begin{equation} \label{eq:4}
f_R R_{ij}-\frac{1}{2}f(R)g_{ij}+\left(g_{ij}\Box-\nabla_i \nabla_j\right)f_R = 8\pi T_{ij}+f_T T_{ij}+ \left[\bar{p} f_T+\frac{1}{2}f(T) \right] g_{ij}.
\end{equation}
Here, $f_R=\frac{\partial f(R)}{\partial R}$, $f_T=\frac{\partial f(T)}{\partial T}$ and $\bar{p}$ be the effective viscous pressure. Three functional forms are suggested as (i) $f(R,T)=R+2f(T)$, (ii) $f(R,T)=f_1(R)+f_2(T)$ and (iii) $f(R,T)=f_1(R)+f_2(R)f_3(T)$. In order to formulate the cosmological model, in this problem, we have assumed $f(R,T)=f_1(R)+f_2(T)$, where $f_1(R)=\mu R$ and $f_2(T)=\mu T$, $\mu$ be the scaling constant. Now, the $f(R,T)$ field equations \eqref{eq:4} reduce to,

\begin{equation} \label{eq:5}
R_{ij}-\frac{1}{2}Rg_{ij}=\left(\frac{8\pi+\mu}{\mu}\right)T_{ij}+\Lambda (T) g_{ij}.
\end{equation}
where, $\Lambda (T)=\bar{p}+\frac{1}{2}T$ can be treated as the effective  cosmological constant that evolves with cosmic time.  \\

We have organised the paper as follow: In section II, the basic equations are formed in an anisotropic space-time along with the dynamical parameters. In section III, the dynamics of the model derived with a variable deceleration parameter and the behaviours are analysed. Section IV contains the results and discussions on the model.

\section{Formation of Basic Equations}
In this section, we have presented the mathematical formalism of the cosmological model to study the dynamical behaviour with an anisotropic space-time. The standard FLRW model is homogeneous and isotropic. The CMB observation has suggested a small amount of anisotropy at the late time of cosmic evolution. So, in order to get into the study of anisotropy, we consider an anisotropic but homogeneous space-time dubbed as Bianchi $VI_h$ space-time. The subscript can take integral values namely  $h=-1,0,1$. In some of our earlier studies \cite{Mishra18b}, we have observed that $h=-1$ provides significant results on the cosmological problems as compared to $h=0,1$, therefore we shall consider Bianchi type $VI_{-1}$ space-time as, 

\begin{equation} \label{eq:6}
ds^2 = dt^2 - C_{1}^{2}dx^2- C^{2}_{2}e^{2x}dy^2 - C^{2}_{3}e^{-2x}dz^2,
\end{equation}
where, $C_{i}=C_i(t), i=1,2,3$. In literature, several cosmological models have been constructed with isotropic space-time and a perfect fluid distribution of the Universe. However, the possibility of the presence of viscosity in the cosmic fluid may be considered. Therefore, here we intend to consider the bulk viscous fluid along with the usual cosmic fluid. It is worth to mention here that the viscous fluid models are instrumental in explaining the early phase cosmic acceleration and the observed highly isotropic matter distribution on the high entropy per baryon. Also, the strong dissipation due to neutrino viscosity may considerably reduce the anisotropy of black- body radiation. There have been considerable interests in cosmological models with bulk viscosity, since bulk viscosity leads to the accelerated expansion phase of the early Universe, popularly known as the inflationary phase \cite{Brevik17a,Brevik17b}. Mishra et al. \cite{Mishra18c} have studied the dynamical properties of the cosmological model with viscous cosmology in $f(R,T)$ gravity. Applying the Misner-Sharp approach, Ahmed and Abbas \cite{Ahmed19} have studied the dissipative gravitational collapse in $f(R, T)$ gravity. Also, Singh and Kumar \cite{Singh19} have studied the holograpic dark energy model in extended modified gravity with the presence of bulk viscosity.  The energy momentum tensor $T_{ij}$ for the viscous fluid can be expressed as
\begin{equation} \label{eq:7}
T_{ij}=(\rho+\bar{p})u_{i} u_{j}- \bar{p} g_{ij},
\end{equation}
where the proper energy density and bulk viscous pressure are respectively  $\rho$ and  $\bar{p}=p-\xi \theta$ with $\xi$ be the bulk viscous coefficient. In the co-moving coordinate system, we have $u^{i}=(0,0,0,1)$; $u^{i}=\delta^{i}_{4}$ satisfying $g_{ij}u^iu^j=1$. The field equations \eqref{eq:5} for the metric \eqref{eq:6} and energy momentum tensor \eqref{eq:7} can be expressed as,

\begin{eqnarray}
\frac{\ddot{C}_2}{C_2}+\frac{\ddot{C}_3}{C_3}+\frac{\dot{C}_2\dot{C}_3}{C_2 C_3}+ \frac{1}{C_1^{2}}&=& -\beta\bar{p}+\frac{\rho}{2},  \label{eq:8}\\
\frac{\ddot{C}_1}{C_1}+\frac{\ddot{C}_3}{C_3}+\frac{\dot{C}_1\dot{C_3}}{C_1 C_3}- \frac{1}{C_1^{2}}&=&-\beta\bar{p}+\frac{\rho}{2},  \label{eq:9}\\
\frac{\ddot{C}_1}{C_1}+\frac{\ddot{C}_2}{C_2}+\frac{\dot{C}_1 \dot{C_2}}{C_1 C_2}- \frac{1}{C_1^{2}}&=&-\beta\bar{p}+\frac{\rho}{2},   \label{eq:10}\\
\frac{\dot{C_1}\dot{C_2}}{C_1 C_2}+\frac{\dot{C_2}\dot{C_3}}{C_2 C_3}+\frac{\dot{C_3}\dot{C_1}}{C_3 C_1}-\frac{1}{C_1^{2}}&=& \beta\rho -\frac{\bar{p}}{2}, \label{eq:11}  \\
\frac{\dot{C_2}}{C_2}-\frac{\dot{C_3}}{C_3}&=&0. \label{eq:12}
\end{eqnarray}

An over dot on the field variable denotes the differentiation with respect to  cosmic time $t$ and $\beta= \left(\frac{8 \pi}{\mu}+\frac{3}{2}\right)$. Due to this non-linearity nature of the field equations \eqref{eq:8}-\eqref{eq:12}, it would be difficult to determine the physical and geometrical parameters. Therefore, we shall express the metric potentials in terms of the directional Hubble parameters $H_j$ along the orthogonal coordinate axes as,  $H_{j}=\frac{\dot{C}_j}{C_j}, j=1,2,3$ such that, the mean Hubble parameter, $H= \dfrac{1}{3} (H_j), j=1,2,3$. We can infer from the field equation \eqref{eq:12}, $H_2=H_3$ by suitably absorbing the integration constant. Also, the Hubble parameter and scale factor can be related as $H=\frac{\mathcal{\dot{R}}}{\mathcal{R}}$. It is to note here that no direct relation is possible between $H_1$ and $H_3$, unless we make a prior assumption. Since our study is basically on an anisotropic cosmological model, apart from the anisotropy incorporated in the space-time, here we shall incorporate some amount of anistropy among the field variable in the form $H_{1} = k H_{3}, k \neq 1$ such that the dynamical behaviour can be studied with additional anisotropy at the background. The field equations can be derived from \eqref{eq:8}-\eqref{eq:12} expressed in term of Hubble parameter as,

\begin{eqnarray}
\left(\frac{6}{k+2}\right)\dot{H}+\left(\frac{27}{k^2+4k+4}\right)H^2+\mathcal{R}^{-\frac{6k}{k+2}}&=&-\beta\bar{p}+\frac{\rho}{2},  \label{eq:13} \\
3\left(\frac{k+1}{k+2}\right)\dot{H}+9\left(\frac{k^2+k+1}{k^2+4k+4}\right)H^2-\mathcal{R}^{-\frac{6k}{k+2}}&=&-\beta\bar{p}+\frac{\rho}{2},  \label{eq:14}\\
9\left(\frac{2k+1}{k^2+4k+4}\right)H^2-\mathcal{R}^{-\frac{6k}{k+2}}&=& \beta \rho-\frac{\bar{p}}{2}. \label{eq:15}
\end{eqnarray}
The above set of field equations still poses some degree of difficulty to provide an explicit solution for the dynamical parameters. In view of this, we wish to adopt some algebraic approaches to express the effective pressure and energy density of the matter as a function of the Hubble parameter. We express the Einstein tensor of \eqref{eq:13}-\eqref{eq:15} with the respective indices as, $S_1(H,k)=\left(\frac{6}{k+2}\right)\dot{H}+\left(\frac{27}{k^2+4k+4}\right)H^2+\mathcal{R}^{-\frac{6k}{k+2}}$, $S_2(H,k)=3\left(\frac{k+1}{k+2}\right)\dot{H}+9\left(\frac{k^2+k+1}{k^2+4k+4}\right)H^2-\mathcal{R}^{-\frac{6k}{k+2}}$ and $S_3(H,k)=9\left(\frac{2k+1}{k^2+4k+4}\right)H^2-\mathcal{R}^{-\frac{6k}{k+2}}$. Now, we can express the effective pressure $\bar{p}$ and energy density $\rho$ as in the following. It is to note here that the  effective pressure that consists of both the proper pressure $p$ and barotropic bulk viscous pressure can be written as, $\bar{p} = p-\xi \theta=p-3\xi H$, where $\xi$ be the coefficient of bulk viscosity. From \eqref{eq:13}- \eqref{eq:15}, we can obtain an expression for $\bar{p}$ and $\rho$ in terms of Hubble parameter as, 

\begin{eqnarray}
\bar{p} &=& p-3\xi H=-\left[S_1(H,k)+S_3(H,k)\right]\left(\frac{2}{1-4\beta^2}\right)+\left[S_2(H,k)\right]\left(\frac{2}{1-2\beta}\right), \label{eq:16}\\
\rho &=&[S_1(H,k)]\left(\frac{2}{1-4\beta^2}\right)-[S_3(H,k)]\left(\frac{4\beta}{1-4\beta^2}\right).\label{eq:17}
\end{eqnarray}
We can also obtain the effective equation of state (EoS) parameter $\omega_{eff}=\frac{\bar{p}}{\rho}$ and the effective cosmological constant $\Lambda$ as,

\begin{eqnarray}
\omega_{eff} &=&-1+\left[\frac{S_2(H,k)-S_3(H,k)}{S_1(H,k)-2\beta S_3(H,k))}\right](1+2\beta),\label{eq:18}\\
\Lambda &=& [S_1(H,k)+S_3(H,k)]\left(\frac{1}{1+2\beta}\right).\label{eq:19}
\end{eqnarray}

\section{Simplified Model Dynamics}
In eqns. \eqref{eq:16}-\eqref{eq:19}, the dynamical parameters of the cosmological models are expressed which will enable us to study the dynamical behaviour of the model. Since the parameters are expressed in terms of Hubble rate, we shall consider a form of the Hubble parameter as  $H=\frac{\dot{\mathcal{R}}}{\mathcal{R}}=\frac{1}{3}\left(a+\frac{b}{t}\right)$ such that the scale factor can be obtained as, $\mathcal{R}=e^{at}t^{b}$ . This scale factor is called hybrid scale factor and the constants  $a,$ $b$ are positive arbitrary constant and can  be calculated from the background cosmology. The values of the parameters $a$ and $b$ have been found in the range $0< [a,b]<1$. These values have been further refined in some earlier research works \cite{Mishra15} where $b$ remains in $[0,\frac{1}{3}]$ and $a>0$ left open. In this proposed scale factor, the deceleration parameter approaches to $-1$ at late time and at the initial stage remain as constant $(\frac{1-b}{b})$. In this range of the model parameters value, we are intending to examine the accelerating behaviour of the model in viscous fluid scenario.  So, we shall express all the dynamical parameters with respect to cosmic time for elaborating its behaviour  during the cosmological evolution. Here, the scale has been fixed such that 1 unit of cosmic time = 10 billion years. However, for comparing the result with recent observational data, the physical parameters of the model can be illustrated graphically with respect to the redshift $z$. This is possible with the defined relation between the scale factor and the redshift as, $z = \dfrac{1}{\mathcal{R}}-1.$ The recent observational results predicts a transition redshift in the range $0.4 \leq z_{t} \leq 0.8$ \cite{Mishra17}; further the Planck collaboration suggested the range as $0.19 \leq z_{t} \leq 0.76$ \cite{planck}. To keep the redshift in the desired range, the  representative values of the parameters are $a = 0.14,$ $ b = 0.32,$ $k = 1.0001633$. In the same note, $\beta=\left(\frac{8\pi}{\mu}+\frac{3}{2}\right)$ and $\xi$ have been analysed within preferred ranges. The value of anisotropic parameter $k$ have been constrained keeping the amount of anisotropy added and the recent anisotropic behavior of Universe at a small scale \cite{Mishra18e}. We consider the generic constants, in Planckian unit system as $(c = G = \hbar = 1)$. Substituting the  form of Hubble parameter, eqns. \eqref{eq:16}-\eqref{eq:17} reduce to, \\
 
\begin{eqnarray}
\bar{p}&=& -\frac{2}{1-4\beta^2}\left[\frac{(k-1)+2\beta(k+1)}{(k+2)}.\left(\frac{b}{t^2}\right)\right] \nonumber \\
&+&\frac{2}{1-4\beta^2}\left[\frac{(k^2-k-3)+2\beta(k^2+k+1)}{(k+2)^2}.\left(a+\frac{b}{t}\right)^2\right] \\
&-&\frac{2}{1-2\beta}.\left({e^{at}t^b}\right)^{-\frac{6k}{k+2}}, \nonumber\\
\rho&=&\frac{2}{1-4\beta^2}\left[-\frac{2}{(k+2)}.\left(\frac{b}{t^2}\right)\right] \nonumber \\
&+&\frac{2}{1-4\beta^2}\left[\frac{3-2\beta(2k+1)}{(k+2)^2}.\left(a+\frac{b}{t}\right)^2\right]\nonumber \\
&+&\frac{2}{1-2\beta}.\left({e^{at}t^b}\right)^{-\frac{6k}{k+2}}. 
\end{eqnarray}

\begin{figure}[h!]
\centering
\includegraphics[width=90mm]{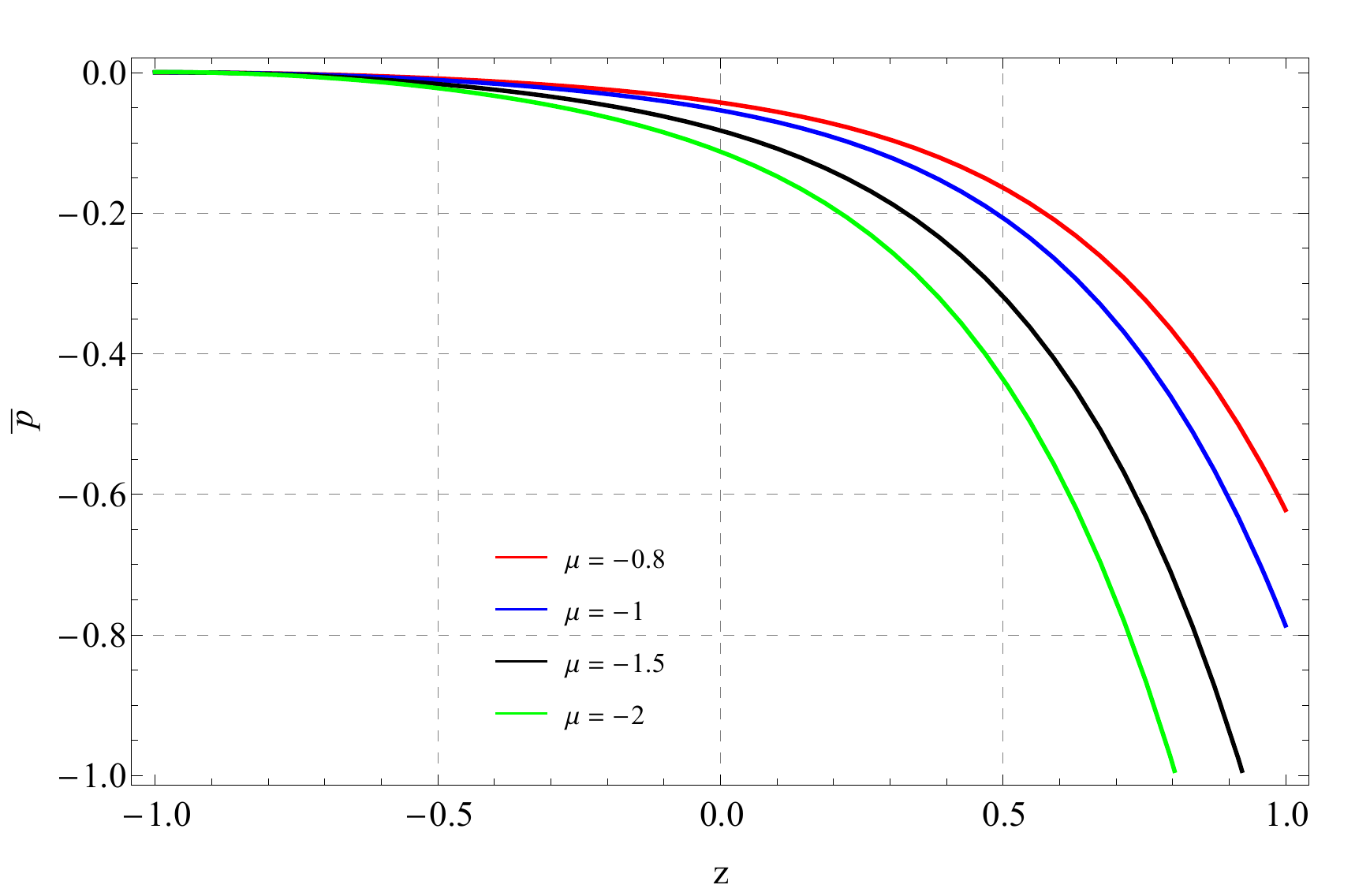}

\hspace{2cm}\caption{Viscous pressure vs Redshift for representative values of $\mu$ }
\end{figure}
\begin{figure}
\centering
\includegraphics[width=90mm]{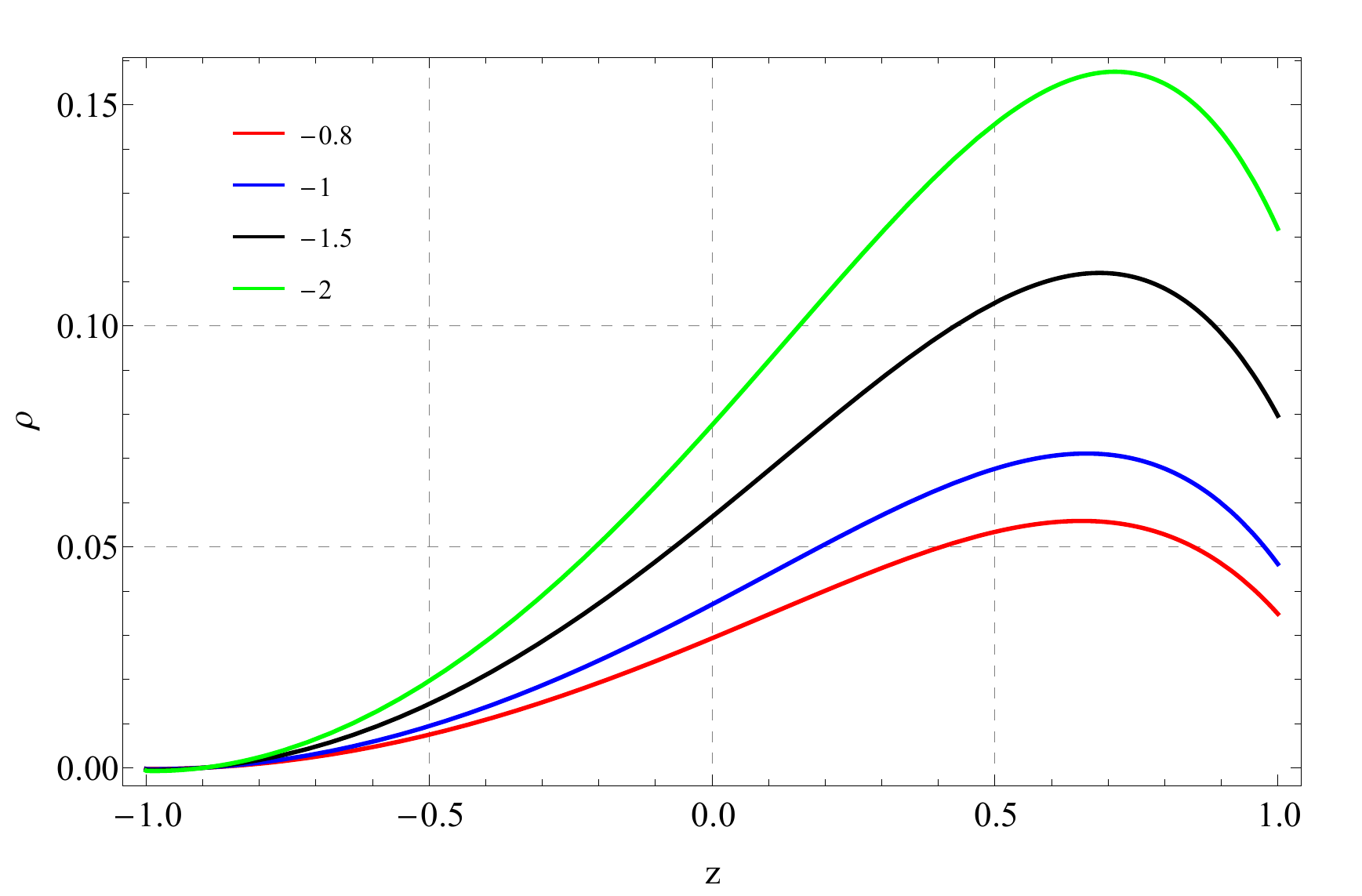}
\caption{Energy density vs Redshift for representative values of $\mu$}
\end{figure}

Fig. 1 shows that the bulk viscous pressure remains negative for the representative values of the scaling constant $(\mu)$ throughout the evolution. The rate of negative pressure is suddenly high in the figure at the initial phase of evolution which may explain the inflationary epoch (exponential expansion of space in the early Universe) that lasted between $10^{-36}$ seconds to $10^{-32}$ seconds after the big bang. Following the inflation period, the Universe continued to expand but at a slower rate, which can be observed from the figure. Eventually, the rate of negative pressure increases exponentially towards late phase of evolution that mimics the accelerated expansion. This sudden increase in negative pressure may sufficiently explain the theory of dark energy that began over $4.82$ billion years ago. The nature of energy density (Fig. 2), as expected, is positive and decreasing. It decreases from a high value at an early time to small values at late time. This observation confirms the fact that the density of matter decreases as the Universe expands because the volume of the space increases. One can also note that with decrease in coupling constant value, the energy density increases in early epoch showing a clear effect of it. However, the effect of $\mu$ vanishes towards the late epoch which can be observed as the different curves in the figure merge together towards $z=-1.$\\

Since the dynamical properties of the Universe are computed through physical parameters with hybrid scale factor, the effective cosmological constant and bulk viscous coefficient can be obtained as follows,

\begin{eqnarray}
\Lambda&=&\frac{2}{(k+2)(1+2\beta)}\left[-\frac{b}{t^2}+\left(a+\frac{b}{t}\right)^2\right], \\
\xi &=& \frac{2}{3(1-4\beta^{2})}\left[ \frac{(2k+1)(\beta-1)+(k^{2}+k+1)(2\beta-0.5)}{(k+2)^{2} \left(bt^{-1}+a \right)^{-1}}+ \frac{b(k^{2}+3k+2)(0.5-2\beta)}{t^{2}\left(bt^{-1}+a \right)(k+2)^{2}}\right] \nonumber \\ 
&+&\frac{2}{3(1-4\beta^{2})}\left[\frac{3(1-2\beta)}{e^{\frac{2at}{k+2}}t^{\frac{2b}{k+2}} \left( bt^{-1}+a\right)} \right].
\end{eqnarray}

\begin{figure}[h!]
\centering
\includegraphics[width=90mm]{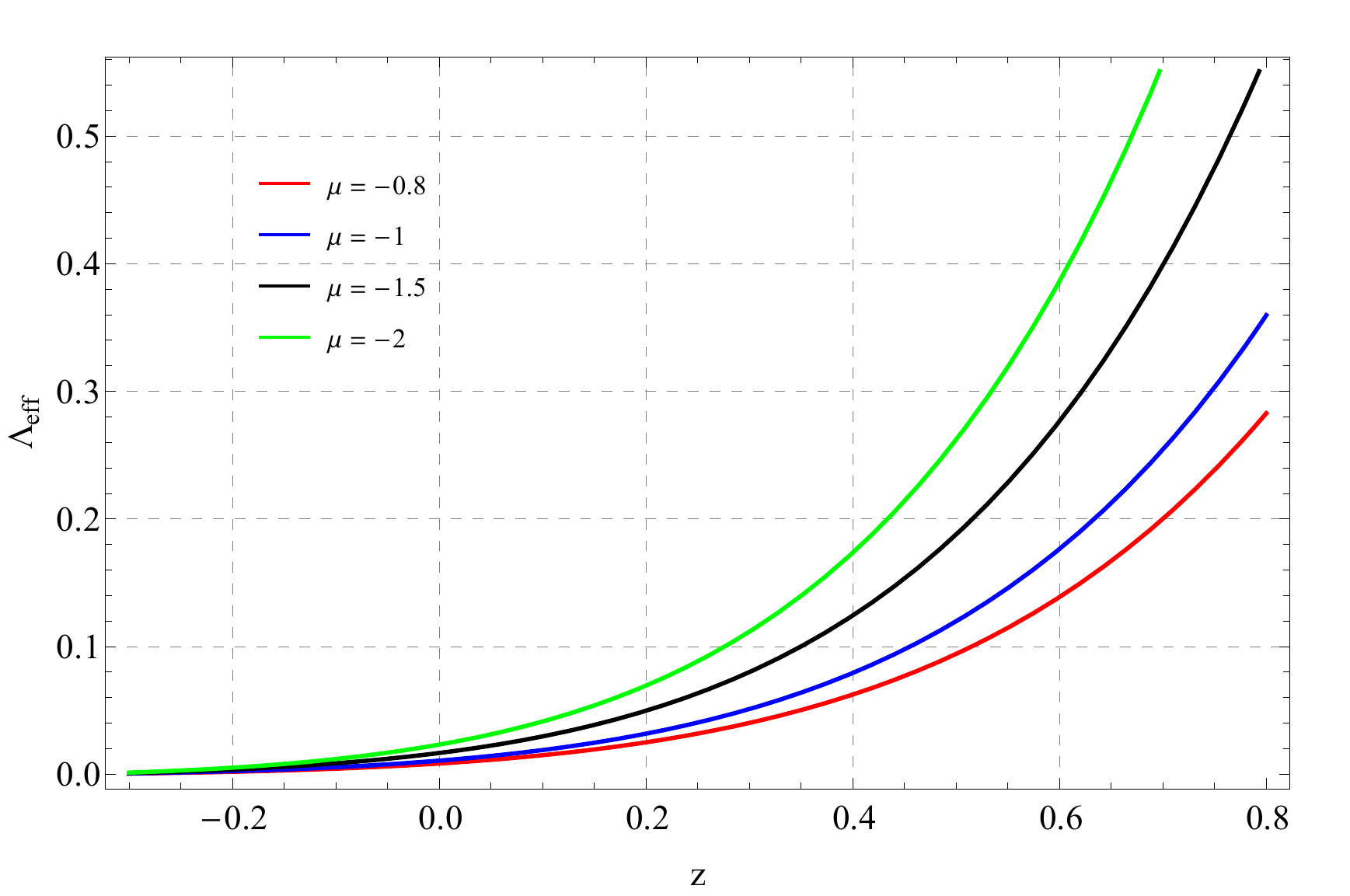}
\caption{Evolution of effective cosmological constant vs Redshift }
\end{figure}
\begin{figure}
\centering
\includegraphics[width=90mm]{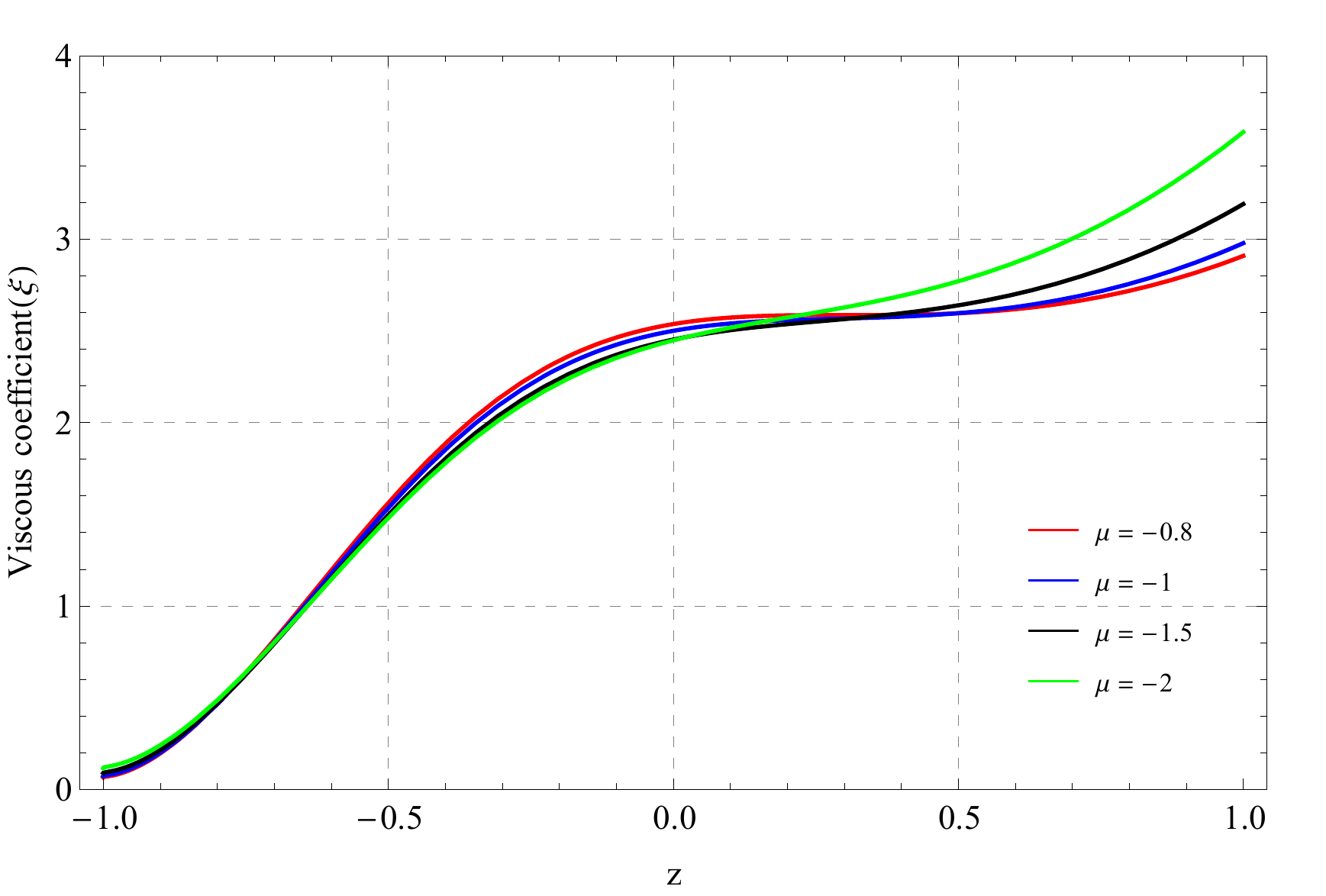}
\caption{Evolution of coefficient of bulk viscous fluid vs Redshift }
\end{figure}

The dynamical variation of the effective cosmological constant for the representative values of $\mu$ has been presented in Fig. 3. It is observed that the evolutionary behaviour of $\Lambda$ is affected by the choice of the value of $\mu$. $\Lambda$ curves with high values of $\mu$ remain below the curves with lower values of $\mu.$ However, at remote past, the effective cosmological constant $\Lambda$ vanishes at late epoch. In fact, as it appears from the figure in general, $\Lambda$ varies from large positive values in early epoch to almost vanishingly small values at late time. This behaviour may describe the phenomenon of late time cosmic acceleration where cosmological constant vanishes. One may note that, the repulsive nature of the cosmological constant is due to a negative pressure which can be considered as bulk viscous pressure for the present model.\\

Fig. 4 shows the evolution of bulk viscous coefficient $(\xi)$ with representative values of the model parameter $\mu.$ For negative values of $\mu,$ the viscous coefficient shows an oscillating nature and gradually attains a vanishingly small value for higher positive values of $\mu.$ The reason behind this nature may be due to some deviation in bulk viscous coefficient. So, one should choose suitable negative values of $\xi$ in order to know its behaviour through the evolution. An important observation in this figure is that the curves are more dominant in the early deceleration phase of evolution but the effect reduces gradually and vanishes at the late epoch. This behavior indicates that, the bulk viscous coefficient plays a vital role at early epoch, possibly providing a strong source for anisotropy.

Consequently, to have an insight in the dynamics of state of matter along with the cosmic evolution, we have derived the EoS parameter for the model as:

\begin{eqnarray}
\omega_{eff}&=&-1+(1+2\beta)\nonumber \\
			&\times&\left[\frac{-(k^2+3k+2)\left(\frac{b}{t^2}\right)+(k^2-k)\left(a+\frac{b}{t}\right)^2}{-2(k+2)\left(\frac{b}{t^2}\right)+3\left(a+\frac{b}{t}\right)^2-2\beta\left[(2k+1)\left(a+\frac{b}{t}\right)^2\right]+(1+2\beta)(k+2)^2\left(e^{at}t^b\right)^{-\frac{6k}{k+2}}}\right].
\end{eqnarray}

\begin{figure}[h!]
\centering
\includegraphics[width=90mm]{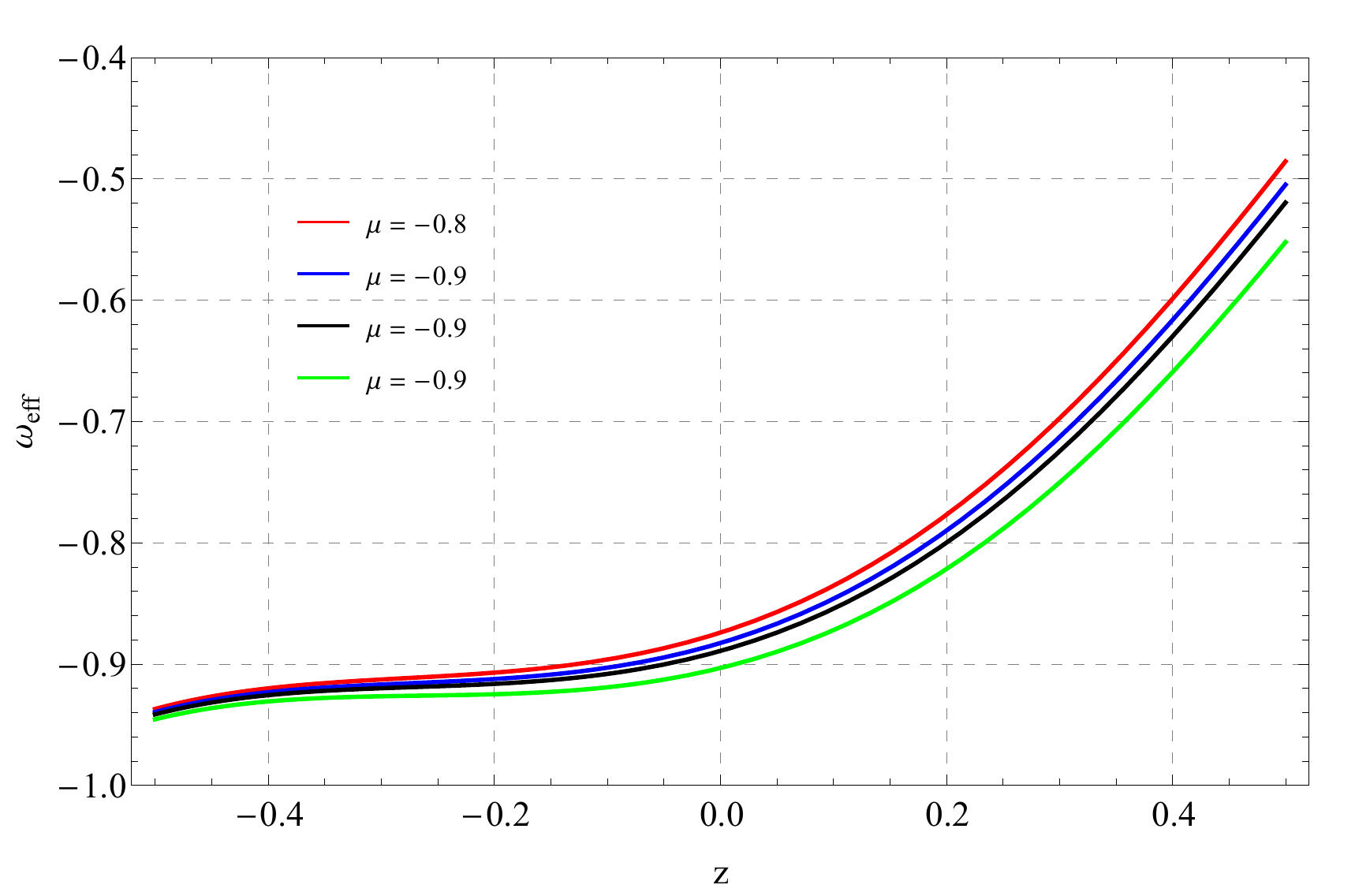}
\caption{Evolution of effective EoS parameter for representative values of $\mu$}
\end{figure}
\begin{figure}
\centering
\includegraphics[width=90mm]{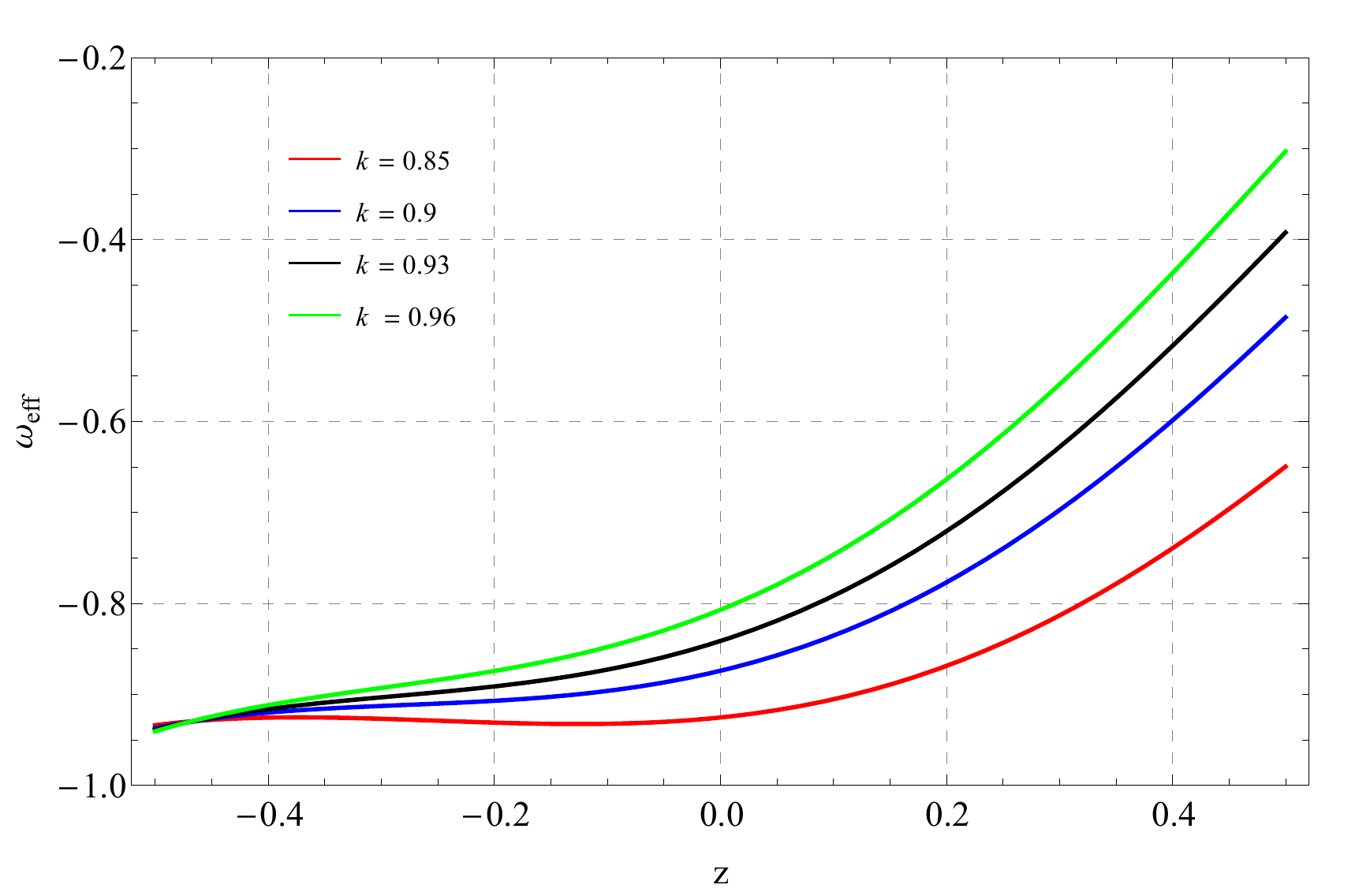}
\caption{Evolution of effective EoS parameter for representative values of anisotropic parameter}
\end{figure}

Fig. 5 shows a concrete dynamical behavior through the effective EoS parameter $(\omega_{eff})$ against the redshift of evolution. While plotting the figure, we considered the free parameters ensuring a positive energy density and negative pressure throughout the cosmic evolution in the model. To keep $\omega_{eff}$ in preferred observational data range, we have incorporated a set of values for the free parameters as discussed earlier in this section.  More precisely, the EoS parameter stays in the quintessence region till late phase. On the same note, we have taken the negative value of the coupling parameter $\mu$ in Fig. 5. It is clear from the figure that for negative $\mu$ values, $\omega_{eff}$ starts evolving almost from a quintessence region $(\omega_{eff} \geq -1)$ during the early phase of evolution. Though the $\omega_{eff}$ curves have started evolving from different regions for different values of $\mu$ but eventually towards late phase of evolution they show same nature falling in the quintessence region as indicated in Fig. 5. It can be noted that when the value of $\mu$ increases, $\omega_{eff}$ increases most rapidly in the initial phase of evolution. Fig. 6 represents the behaviour of the EoS parameter for the representative value of the anisotropic parameter. It has been observed that for a smaller $k$ value the evolution starts from small negative value and increases with the increase in the value of $k$. But in spite of different values of the anisotropic parameter at late phase all merged and stay in the range $[-0.88,-0.86]$.

\begin{figure}[h!]
\minipage{0.90\textwidth}
\centering
\includegraphics[width=95mm]{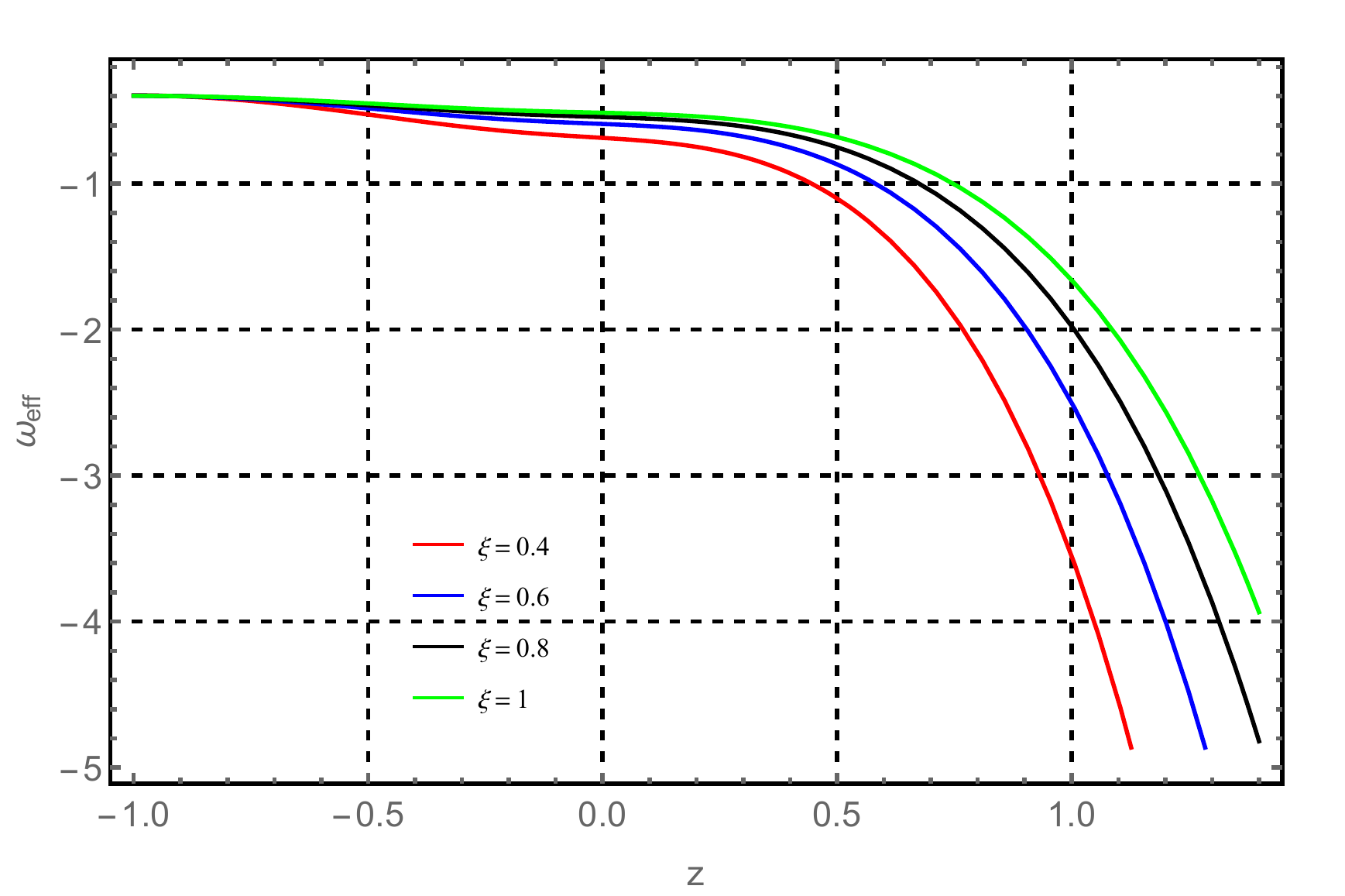}
\caption{Evolution of effective EoS parameter for representative values of viscous coefficient $\xi$}
\endminipage
\end{figure}

We have observed in Fig. 4 that the viscous coefficient gradually decreases from higher positive value and vanishes at late time. In addition, we are interested to assess the behaviour of EoS parameter with some representative values of the viscous coefficient as shown graphically in Fig. 7. The observation is that unlike for the parameters $a,$ $b$ and $k$, the presence of viscous coefficient strongly affect $\omega_{eff}$ at early time of evolution. In order to investigate this effect of viscous coefficient on the prescribed model, we have considered representative values of $\xi$ in the range $[0,1].$ The viscous coefficient behaves same outside the range in a larger scale approach.  It is evident from the figure that the presence of viscous fluid incorporates a substantial amount of anisotropy and affect the early phase of cosmic evolution. This can be observed by the increasing trend of $\omega_{eff}$ for different $\xi$ values as we move back to past. The effective EoS is affected by deviation in viscous coefficient at late phase of evolution which can be observed by smoothly merging of all four lines for different values of $\xi$ (Fig. 7). This nature is also obvious in Fig. 1, Fig. 2, Fig. 3, Fig. 4 and Fig. 6 which concretes the facts that at early phase, the viscous coefficient has a substantial contribution to energy density as well as viscous pressure for which the dynamics of effective EoS is greatly affected. The reason behind nature of effective EoS parameter in the late phase of expansion may be due to the presence of strongly dominant dark energy over the presence of viscous coefficient \cite{Mishra18e,Mishra17}. One can also note that, for different values of viscous coefficient in the preferred range [0,1], $\omega_{eff}$ decreases smoothly and lies in the quintessence region as mentioned by recent Planck collaboration data \cite{planck}. Hence, the present model behaves like quintessence field.\\

Next, we shall examine the behaviour of the energy conditions of the model as the extended gravity requires the violation of strong energy condition at least at the late epoch of the cosmic evolution. As standard matter is assumed to satisfy the necessary energy conditions, so for a viscous fluid distribution in $f(R,T)$ gravity, the energy conditions can be expressed as,
Null Energy Condition(NEC): $\rho+\bar{p} \geq 0$; Weak Energy Condition(WEC): $\rho+\bar{p} \geq 0,$ $\rho \geq 0$; Strong Energy Condition(SEC): $\rho+3\bar{p}\geq0$ and Dominant Energy Condition(DEC): $\rho- \bar{p}\geq0,$ $\rho\geq0$. In this model, the energy conditions can be derived as, 
\begin{eqnarray}
\rho+\bar{p}&=&\frac{2}{1-4\beta^{2}}\left[\left( \frac{(k+1)(2\beta-1)}{k+2}\right) \left( \frac{b}{t^{2}}\right)+\left( \frac{(k^{2}-k)(2\beta+1)}{(k+2)^{2}}\right) \left(a+\frac{b}{t} \right)^{2}\right],\\
\rho-\bar{p}&=&\frac{2}{1-4\beta^{2}}\left[ \left( \frac{k-3+2\beta(k+1)}{k+2}\right)\left( \frac{b}{t^{2}}\right)+\left(\frac{(-k^{2}+k-6)-2\beta(k^{2}+3k+2)}{(k+2)^{2}} \right) \left(a+\frac{b}{t} \right)^{2} \right], \nonumber\\
&+&\frac{4}{1-2\beta}\left( e^{at}t^{b}\right)^{\frac{-6k}{k+2}}\\
\rho+3\bar{p}&=&\frac{2}{1-4\beta^{2}}\left[ \left( \frac{-3(k-1)-6\beta(k+1)}{k+2}\right)\left( \frac{b}{t^{2}}\right)+\left( \frac{3(k^{2}-k-2)+2\beta(3k^{2}+k-2)}{(k+2)^{2}}\right)\right] \nonumber\\ &-&\frac{4}{1-2\beta} \left( e^{at}t^{b}\right)^{\frac{-6k}{k+2}}.  
\end{eqnarray}

The energy conditions, in Fig. 8, are observed to change dynamically with the cosmic evolution. The SEC is satisfied in early phase of evolution upto $t=0.55$ Gyr and afterwards, there is a clear violation of this condition (green line). The blue curve shows that DEC is satisfied throughout the evolution. Moreover, the WEC and NEC are satisfied in the model. The characteristics of the energy conditions might be due to the presence of viscous pressure in the matter field.
Also, it can be inferred that the present model in the framework of extended gravity favours an accelerated expansion of the Universe, hence the violation of SEC has become inevitable and the same has been shown in Fig. 8.

\begin{figure}[h!]
\minipage{0.90\textwidth}
\centering
\includegraphics[width=95mm]{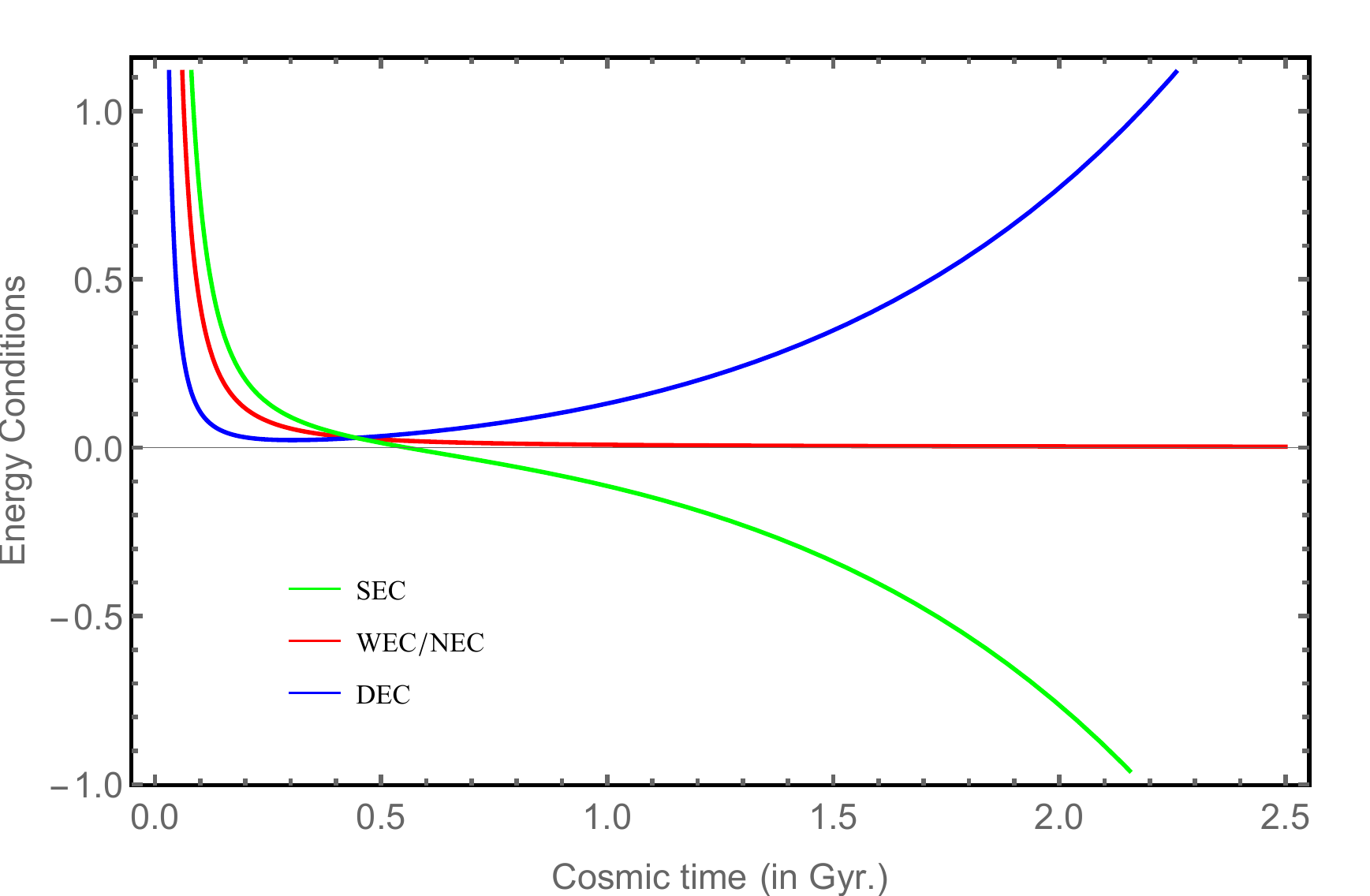}
\caption{Energy conditions vs cosmic time in Gyrs.}
\endminipage
\end{figure}

\section{Results and Discussions} 

The cosmological model of the Universe in an extended theory of gravity has been presented with an anisotropic space-time. A dissipative cosmic fluid in the form of bulk viscosity is chosen to study its effect on the cosmic dynamics particularly the equation of state parameter. A mathematical formalism is discussed for framing the cosmological model, where we obtain a time varying cosmological constant as an effect of the modification of geometry in the action. The viscous pressure, energy density, EoS parameter, viscous coefficient and effective cosmological constant are calculated with the developed mathematical formalism and its graphical representations are given. Throughout this work the model parameter $\mu$ plays a key role to describe an anisotropic acceleration Universe. It is observed that, the choice of the extended gravity parameter $\beta$ affects the behaviour of the effective cosmological constant $(\Lambda)$ (Fig.3). $\Lambda$ varies from large positive values with high value of $\mu$ at early epoch and remains below with lower value of $\mu$ at late epoch. The bulk viscous coefficient initially increases from large negative values and after attending a peak decreases to a vanishingly small value(Fig.4). The extended gravity parameter $\mu$ decreases the requirement of bulk viscous fluid. The dynamics of the Universe in the form of the evolution of the EoS parameter is discussed in Fig.5 and Fig.6. We have shown the behaviour both with the representative values of the model parameter and anisotropic parameter. We obtained the EoS parameter as predicted from the negative choice of $\beta$, lying in the range of $-0.9$ to $-1$ at the present epoch. Similarly, the effect of viscous coefficient has shown through  the evolution of EoS parameter(Fig.7). We have obtained that the presence of viscous fluid incorporates a substantial amount of anisotropy and affect at early stage of cosmic evolution. With different values of viscous coefficient$(\xi)$, $\omega_{eff}$ increases slowly and lies in the quintessence region in the cosmic evolution process for different $\xi$ values. \\ 

The present investigation provides some useful insight into the cosmic dynamics and evolution of the EoS parameter in presence of a dissipative fluid. Our model is quite compatible with recent observational result that favours the accelerated expansion of the Universe. However, we have used a very specific form of the functional $f(R,T)$. A more general form of the functional $f(R,T)$ in place of the one used in the present work, may provide some more information about the viscous cosmic dynamics. 

\section*{Acknowledgements}
ST acknowledges Rashtriya Uchchatar Shiksha Abhiyan (RUSA), Ministry of HRD, Govt. of India for the financial support. BM and SKT acknowledges Inter-University Center for Astronomy and Astrophysics (IUCAA), Pune, India for hospitality and support during an academic visit where a part of this work is accomplished. The authors are thankful to the honourable referee for the valuable suggestions and comments for the improvement of the paper.


\begin{thebibliography}{99}

\bibitem{Reiss98} A.G. Riess et al., \textit{Astron J.}, \textbf{116}, 1009 (1998).
\bibitem{Reiss99} A.G. Riess et al.,  \textit{Astron J.}, \textbf{117}, 707 (1999).
\bibitem{Spergel03} D.N. Spergel et al., \textit{Astrophys. J. Suppl.}, \textbf{148}, 175  (2003).
\bibitem{Abaz24} K. Abazajian,  \textit{Astron. J.}, \textbf{128}, 502 (2004).
\bibitem{Pope24} A. C. Pope et al., \textit{Astrophys. J.}, \textbf{607}, 655 (2004).
\bibitem{Noji00} S. Nojiri, S.D. Odintsov, \textit{Int. J. Geom. Meth. Mod. Phys.} \textbf{4}, 115-146, (2007).
\bibitem{Noji11} S. Nojiri, S.D. Odintsov, \textit{Phys. Rept.} \textbf{505}, 59-144, (2011).
\bibitem{Capo} S. Capozziello, M. De Laurentis, \textit{Phys. Rept.} \textbf{509}, 167-321,(2011).

\bibitem{Bamba12} K. Bamba et al., \textit{Astrophys Space Sci.}, \textbf{342}, 155 (2012).
\bibitem{Harko11} T. Harko, F.S.N. Lobo, S. Nojiri, S.D. Odintsov, \textit{Phys. Rev.D} \textbf{84},024020 (2011).
\bibitem{Das16} A. Das, F. Rahaman, B.K. Guha, S. Ray, \textit{Eur. Phys. J. C}, \textbf{76}, 654 (2016).
\bibitem{Deb18} D. Deb, F. Rahaman, S. Ray, B.K. Guha, \textit{JCAP}, \textbf{03}, 44 (2018).
\bibitem{Fisher18} S. B. Fisher, E. D. Carlson, \textit{Phys. Rev. D}, \textbf{100}, 064059 (2019).
\bibitem{Shabani18} H. Shabani, A.H. Ziaie, \textit{Eur. Phys. J. C}, \textbf{78}, 397 (2018). 
\bibitem{Tripathy19} S.K. Tripathy, R.K. Khuntia, P. Parida, \textit{Eur. Phys. J. Plus}, \textbf{134}, 504 (2019).
\bibitem{Barbar20} A. H. Barbar, A. M. Awad, M. T. AlFiky, \textit{Phys. Rev. D}, \textbf{101}, 044058 (2020). 
\bibitem{Wu18} J. Wu, G. Li, T. Harko, S.D. Lang, \textit{Eur. Phys. J. C}, \textbf{78}, 430 (2018).
\bibitem{Elizalde19} E. Elizalde, M. Khurshudyan, \textit{Int. J. Mod. Phys. D}, \textbf{28}, 1950172 (2019). 
\bibitem{Khan18} S. Khan, M. S. Khan, A. Ali, \textit{Mod. Phys. Lett. A}, \textbf{33}, 1850065.
\bibitem{Khan19} M. S. Khan, S. Khan, \textit{Gem Relativ. Grav.}, \textbf{51}, 148 (2019).
\bibitem{Tripathy20} S.K. Tripathy, B. Mishra, \textit{Chin. J. Phys.}, \textbf{63}, 448 (2020).
\bibitem{Mishra20} B. Mishra, S.K. Tripathy, \textit{Phys. Scr.}, \textbf{95}, 095004 (2020).
\bibitem{Tripathy2020} S. K. Tripathy et al., \textit{Phys. Scr.}, \textbf{95}, 115001 (2020).
\bibitem{Saridakis20} E. N. Saridakis, S. Myrzakul, K. Myrzakulov, K. Yerzhanov, \textit{Phys. Rev. D}, \textbf{102}, 023525 (2020).

\bibitem{Barrow1977} J. D. Barrow, R. A. Matzner, \textit{Mon. Not. Roy. Astron. Soc.}, \textbf{181}, 719 (1977).

\bibitem{Pavon1993}D. Pavon, W. Zimdahl, \textit{Phys. Lett. A}, \textbf{179}, 261 (1993).

\bibitem{Padmanabhan87} T. Padmanabhan and S. Chitre: \textit{Phys. Lett. A}, \textbf{120}, 433(1987).

\bibitem{Noureen} I. Noureen, Usman-ul-Haq and S. A. Mardan, \textit{Int. J. Mod. Phys. D} \textbf{30} 04, 2150027 (2021).
\bibitem{CP} C. P. Singh, P. Kumar: \textit{Eur. Phys. J. C} \textbf{74},3070(2014). 
\bibitem{S}Sezgin Aygun, \textit{Turk J. Phys.} \textbf{41}, 436 – 446(2017).
\bibitem{H} H Azmat, M. Zubair and I Noureen, \textit{Int. J. Mod. Phys. D} \textbf{27}, 01, 1750181 (2018).
\bibitem{RP} R Prasad, L K Gupta, G K Goswami and A K Yadav:, Pramana  J. Phys. \textbf{94},135(2020).

\bibitem{Debnath2019} P. S. Debnath, \textit{Int. J. Geom. Methods. in Mod. Phys.}, \textbf{16}, 1950005 (2019).

\bibitem{Odintsov2020} S. D. Odintsov, D. S. Gomez and G. S. Sharov, \textit{Phys. Rev. D}, \textbf{101}, 044010 (2020).

\bibitem{Mishra18b} B. Mishra, S. Tarai, S.K. Tripathy, \textit{Mod. Phys. Lett. A}, \textbf{33}, 1850052 (2018).

\bibitem{Brevik17a} I. Brevik, A.V. Timoshkin, \textit{Int. J. Geom. Meth. Mod. Phys.}, \textbf{14}, 1750061, (2017).
\bibitem{Brevik17b} I. Brevik et al., \textit{Int. J. Mod. Phys. D}, \textbf{26}, 1730024 (2017).
\bibitem{Mishra18c} B. Mishra, S. Tarai, S.K.J. Pacif, \textit{Int. J. Geom. Meth. Mod. Phys.}, \textbf{15}, 1850036 (2018).
\bibitem{Ahmed19} R. Ahmed, G. Abbas, \textit{Can. J. Phys.}, \textbf{97}, 994 (2019).
\bibitem{Singh19} C.P.Singh, V. Kumar, \textit{Gravit.  Cosmol.}, \textbf{25}, 58 (2019).
\bibitem{Mishra15} B. Mishra, S. K. Tripathy, \textit{Mod. Phys. Lett. A}, \textbf{30}, 1550175 (2015).
\bibitem{planck} Planck collaboration, \textit{Astron. and Astrophys.}, \textbf{641}, A6, (2020).
\bibitem{Mishra17} B. Mishra, Pratik P. Ray, S.K.J. Pacif, \textit{Eur. Phys. J. Plus}, \textbf{132}, 429 (2017).
\bibitem{Mishra18e} B. Mishra, S.K. Tripathy, Pratik P. Ray, \textit{Astrophys. Space Sci.},\textbf{363}, 86 (2018).
\end{thebibliography}
\end{document}